\documentclass[journal]{IEEEtran}

\usepackage{amsfonts}
\usepackage{amsmath,accents}

\usepackage[usenames, dvipsnames]{color}
\usepackage{enumerate}

\usepackage{soul} 

\usepackage{cite}

\ifCLASSINFOpdf
  \usepackage[pdftex]{graphicx}
  \DeclareGraphicsExtensions{.pdf,.jpeg,.png}
\else
\fi
\usepackage{amsmath}
\usepackage[linesnumbered,ruled]{algorithm2e}

\usepackage{array}

\ifCLASSOPTIONcompsoc
  \usepackage[caption=false,font=normalsize,labelfont=sf,textfont=sf]{subfig}
\else
  \usepackage[caption=false,font=footnotesize]{subfig}
\fi
\usepackage{url}

\usepackage{pifont}
\newcommand{\xmark}{\ding{55}}%
\newcommand{\cmark}{\ding{51}}%
\usepackage[table]{xcolor}

\begin{document}
%
\title{ActionXPose: A Novel 2D Multi-view Pose-based Algorithm for Real-time Human Action Recognition}


\author{Federico Angelini,~\IEEEmembership{Student Member,~IEEE,}
        Zeyu Fu,~\IEEEmembership{Student Member,~IEEE,}
		Yang Long, ~\IEEEmembership{Senior Member,~IEEE}, \\
		Ling Shao, ~\IEEEmembership{Senior Member,~IEEE},
        and~Syed~Mohsen~Naqvi,~\IEEEmembership{Senior Member,~IEEE}
}

\maketitle

\begin{abstract}
We present ActionXPose, a novel 2D pose-based algorithm for posture-level Human Action Recognition (HAR). The proposed approach exploits 2D human poses provided by OpenPose detector from RGB videos. ActionXPose aims to process poses data to be provided to a Long Short-Term Memory Neural Network and to a 1D Convolutional Neural Network, which solve the classification problem. ActionXPose is one of the first algorithms that exploits 2D human poses for HAR. The algorithm has real-time performance and it is robust to camera movings, subject proximity changes, viewpoint changes, subject appearance changes and provide high generalization degree. In fact, extensive simulations show that ActionXPose can be successfully trained using different datasets at once. State-of-the-art performance on popular datasets for posture-related HAR problems (i3DPost, KTH) are provided and results are compared with those obtained by other methods, including the selected ActionXPose baseline. Moreover, we also proposed two novel datasets called MPOSE and ISLD recorded in our Intelligent Sensing Lab\footnote{Datasets and algorithm will be available at http://www.intellsensing.com/\\research/information-processing/multimodal-surveillance/}, to show ActionXPose generalization performance.
Highly accurate results on MPOSE and ISLD datasets are provided and compared with those obtained by the method baseline, to justify the efficacy of the proposed method.
\end{abstract}


%
\IEEEpeerreviewmaketitle

\section{Introduction}
\label{sec:introduction}
%
%
%
%

\IEEEPARstart{H}{uman} Action Recognition (HAR) is one of the most challenging problems for an Artificial Intelligence (AI) \cite{Herath2017}. It consists of training an AI model to recognise a class of human actions. Depending on the applications, input data can be very different such as conventional RGB videos, infrared data, time-of-flight-based or structured-light-based data (depth data) \cite{Zhang2017a}.

HAR is a crucial task in several applications such as surveillance, cybernetics, human-machine interaction, automated assisted living systems, automated security, autonomous vehicles, gaming and sport analysis. In this paper, our focus is on surveillance.

In this paper, ActionXPose algorithms are presented, as effective multi-view body pose-based methods for HAR exploiting the OpenPose detector \cite{Cao2016}.  

Human actions can be thought as a \emph{stratified} phenomenon \cite{Vishwakarma2013}. Actions such as \emph{standing}, \emph{sitting}, \emph{raised-left-arms-while-sitting} are only \emph{postures} that can be shared between more meaningful actions like \emph{running}, \emph{working} or \emph{carrying-bag}. Such postures are examples of superficial but important information that can be extracted from data by using body pose detectors. In this work, we focus on posture-level HAR, being interested in exploring and leveraging target body limbs positions only. This problem is affected by the well-known challenges in HAR such as intra-class variation, viewpoint changes, target detection and localisation, subject appearance changes, occlusion, self-occlusions and camera movements \cite{Zhang2017a}.

\begin{figure}
\centering
\includegraphics[scale=0.51]{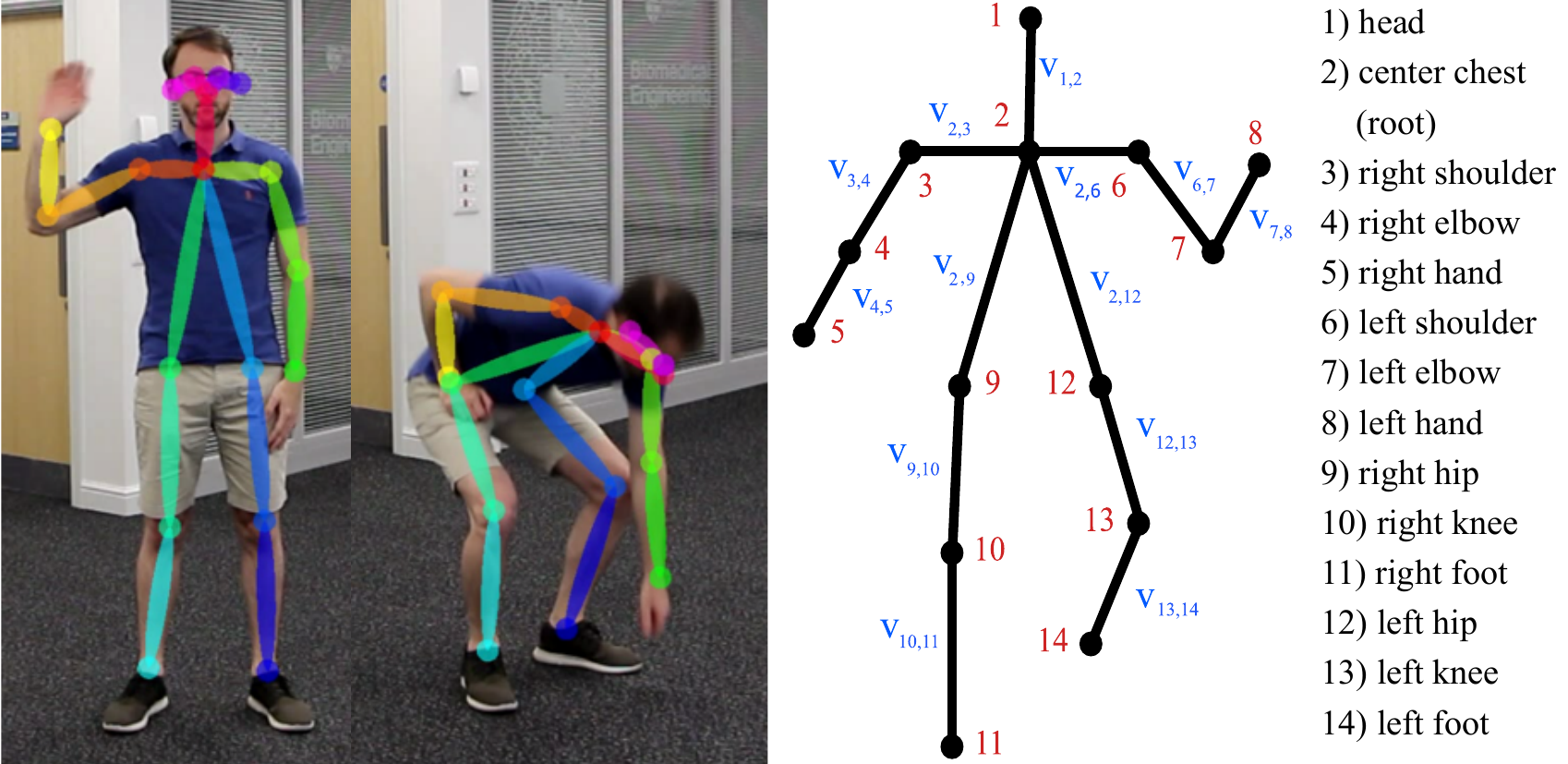}
\caption{Examples of human poses. The pose is determined by detected 2D landmarks, which can be considered as a short and not redundant description of the target posture. Pose detection algorithm is provided in \cite{Cao2016}.}
\label{fig:humanpose}
\end{figure}

Considering posture-level HAR, a number of suitable datasets have been published such as i3DPost \cite{Gkalelis2009b}, Weizmann \cite{Gorelick2007}, IXMAS \cite{Weinland2006x} and KTH \cite{Schuldt2004}. In these datasets, more emphasis is given to \emph{body postures} rather than to the \emph{context} or to unconstrained \emph{recording conditions}. Therefore, these datasets are more suitable when researchers do not consider neither contextual information, i.e. scenario details and objects involved, nor moving camera, strongly cluttered background and occlusions. Conversely, in order to address such challenging problems, suitable datasets have been proposed, such as UCF-101 \cite{Soomro2012}, HMDB-51 \cite{Kuehne2011} and ActivityNet \cite{caba2015activitynet}. In fact, these are large-scale datasets for human activity recognition and understanding. They incorporate huge variability in terms of a wide range of characteristics, such as viewpoint, intra-class action variations, context, background, number of human targets and targets proximity. To the best of our knowledge, nowadays only Convolutional Neural Network (CNN) based methods such as Two-stream \cite{Simoniyan2014, Ma2018} can effectively address such challenging datasets, leveraging the impressive efficacy of CNNs in dealing with colour based features extraction. On the other hand, CNN based methods suffers from \emph{overfitting} problems when trained on \emph{smaller datasets}, such as the posture-level related above mentioned datasets. For example, in the case of surveillance, where camera settings are usually controlled and available datasets are relatively small in comparison with for example ActivityNet, barebones CNN based method could not be effective enough. Furthermore, our experience using OpenPose shows that this method still need further training on bigger datasets to be able to effectively address the unrestricted camera conditions used for example by ActivityNet or UCF-101. Hence, in this paper, the major focus in on posture-related datasets, i.e. i3DPost, Weizmann, IXMAS and KTH. 

Proposed method, ActionXPose, is presented as a development of its baseline counterpart, i.e. OpenPose-baseline. In particular, two versions of ActionXPose have been reported, namely \emph{ActionXPose-basic} and \emph{ActionXPose-advanced}. Regarding the baseline, OpenPose-baseline consists of processing raw data provided by OpenPose without any additional refinements.
In the rest of the paper, we will simply refer to ActionXPose when no distinction is required.

Regarding performance comparisons, ActionXPose results are compared with available state-of-the-art results and also compared with results obtained by OpenPose-baseline. In particular, among available public datasets, in this paper two additional dataset are proposed, MPOSE and ISLD, to measure ActionXPose generalisation abilities.


The proposed work relevance is threefold. First, to the best of our knowledge, ActionXPose represents one of the first attempts of exploiting human 2D poses for HAR. Second, it lays foundation for a deeper and more general approach involving CNN features data alongside human poses for a deeper understanding of human actions. 
Third, ActionXPose provides a \emph{general} learning process where multiple datasets can be combined together in a common framework. Such ability has been proven considering proposed MPOSE and ISLD datasets. 

In short, this paper contribution is twofold:
\begin{enumerate}
	\item A new general HAR algorithm has been specifically designed for posture-level human action recognition which only exploits RGB data for body-pose extraction;
	\item Two new datasets for posture-related HAR problems, MPOSE and ISLD, have been developed providing detected human poses and RGB videos;
\end{enumerate}

Remarkable ActionXPose advantages can be summarised as follows:
\begin{itemize}
	\item \emph{Robustness to camera moving and different subject camera proximity}. In other words, ActionXPose can detect target actions regardless of camera and subject movings in the scene;
	\item \emph{Robustness to viewpoint changes}. In other words, ActionXPose can effectively detect target actions across several point of views;
	\item \emph{High degree of generalization}, due to the fact that ActionXPose detects the action by using human poses rather than RGB information; thus, training data can come from datasets with different resolutions, colours and backgrounds;
	\item \emph{Real-time performance}. In fact, both OpenPose and proposed work can run in real-time, for a fast HAR in surveillance scenarios. 
\end{itemize}
 

The rest of this paper is organised as follows. In the next section, related works have been summarised. In Section \ref{sec:preliminaries}, useful notations and the motivating OpenPose-baseline algorithm are introduced. In Section \ref{sec:ActionXPose}, we discussed technical details about ActionXPose implementation. In Section \ref{sec:mpose_isld_dataset}, details about the proposed MPOSE and ISLD datasets are provided. In Section \ref{sec:experiments}, results of extensive simulations on public and proposed datasets are presented. Conclusions and future works have been discussed in Section \ref{sec:conclusions}.

\section{Related work}
\label{sec:related_work}
State-of-the-art HAR algorithms can be mainly divided into three categories \cite{Herath2017}:
\begin{itemize}
	\item \emph{Hand-crafted features based approaches}. Algorithms in this category show high degree of human participation in the learning process. In other words, human insights are strongly used to solve HAR problems while machine abilities are not fully exploited and often limited to conventional machine learning tasks. Notable approaches rely for example on background subtraction methods \cite{Barnich2015a}, 3D-Histogram of oriented gradients (3D-HOG) \cite{Angelini2018icassp} and Local binary patterns (LPB) features \cite{Zhang2017a, Vishwakarma2013}.
	\item \emph{Deep-learning based approaches}. Algorithms belonging to this category exploit deep neural networks to leave the AI free to explore data without any or with very limited human insights. Notable approaches are based on multiple CNNs, generative models, 3D-CNN and Recurrent Neural Networks (RNN) \cite{Wang2017}. 
	\item \emph{Hybrid approaches}. Algorithms in this category attempt to combine the most promising results from both hand-crafted and deep learning based approaches, researching a convenient trade-off between them \cite{Herath2017}.
\end{itemize}

According to this taxonomy, the method presented in this paper, ActionXPose, is a fully fledged hybrid algorithm to perform posture-level HAR. In fact, ActionXPose is based on \emph{2D human poses} data (Fig. \ref{fig:humanpose}), provided by the human body-pose detector OpenPose \cite{Cao2016}. 
As we will discuss in Section \ref{sec:ActionXPose}, ActionXPose is articulated on a combination of deep learning based algorithms such as Long Short-Term Memory (LSTM) recurrent neural networks  \cite{Hochreiter1997} and Self-Organizing Map (SOM) \cite{kohonen-self-organizing-maps-2001}. The process is also supported by hand-crafted features specifically designed for ActionXPose as well as by well known simple tools such as Principal Component Analysis (PCA). 

Human \emph{skeleton} based approaches are strongly related to the proposed method. As opposite to human \emph{pose} data, skeletal data consists of estimated body landmarks 3D coordinates provided mostly by Kinect camera. Kinect is a structured-infrared-light camera able to reconstruct 3D models of the target area, providing skeleton data for humans in the scene as 3D coordinates. Despite the great advantages provided by 3D data \cite{Liang2015}, Kinect presents numerous limitations. For example, it does not work well in outdoor environments and has a very limited range, which affects its implementation in the surveillance scenarios or in any outdoor applications \cite{Langmann2012}. However, rich literature about how to exploit Kinect data have been published as in most cases skeleton data are descriptive enough to solve the selected HAR problem \cite{Wang2017}. 

Alongside skeleton data, RNN and LSTM have also been exploited in many recent papers, achieving promising results in HAR. In \cite{Liu2017b}, the authors focused skeleton-based action recognition using LSTM networks. In that work, major focus was on implementing trusting gates on the LSTM architecture to allow better action representation.
In \cite{YongDu2015}, a hierarchical RNN approach was implemented to focus on several skeleton subparts, in order to better discriminate which body sub-part is relevant to the performed action. In \cite{Veeriah2015}, authors focused on proposing new gates strategies for LSTM, in order to emphasise salient motion from learning data. In \cite{Zhu2016}, a new regularization term for LSTM has been used in order to learn co-occurrences within skeleton data.
Mainly, all these methods focused on modifying LSTM architectures to better fit 3D skeleton data processing. As opposite, proposed work considers 2D body landmarks coordinates, i.e. human poses, and a pre-defined LSTM architecture. Thus, the major focus is on providing to the LSTM more meaningful input sequences rather than raw data.

Motivated by Kinect limitations, researchers started developing analogues techniques that can provide similar output data using a simple RGB camera. Therefore, highly promising body pose detectors have been published in the last few years, such as DeeperCut \cite{Insafutdinov2016} and OpenPose \cite{Cao2016}. In particular, OpenPose achieves the best performance, opening new research scenarios in HAR. For example, in \cite{Yan2018}, authors have implemented a CNN based method to process 2D poses data provided by OpenPose for HAR, achieving promising results. The authors processed body poses data by using graph convolutional networks. This approach considers temporal information alongside spatial information. However, such method requires a fixed number of frames for each action sample in order to build the action graph, which may affect system flexibility. As opposite, our work focuses on exploiting LSTM networks, which allows full flexibility with respect to space and time, being specifically designed to deal with multivariate temporal sequences with no restrictions on the number of time steps. Moreover, our approach also suggests straightforward solutions to implement additional data exploration levels in the same LSTM framework.
\section{Preliminaries}
\label{sec:preliminaries}
\subsection{Problem's Statement and Notations}
\label{sec:problem_statement}
Let $\mathbb{D} = \{s_i,l_i,w_i\}_{i=1}^{N}$ be the action dataset containing $N$ samples. $s_i$ represents the $i$th subject video (RGB data). $l_i \in \mathcal{L}$ represents the $i$th action label, where $\mathcal{L}$ is the set of target actions and $w_i \in \mathcal{W}$ represents the $i$th viewpoint label, where $\mathcal{W}$ is the set of considered viewpoints. Let $\mathbb{T} \subset \mathbb{D}$ be the chosen training subset and $\mathbb{T}^* = \mathbb{D} \backslash \mathbb{T}$ the testing subset. 

The pose detector provides a mapping between RGB data to body landmarks as: 
\begin{equation}
	s_i(t) \xrightarrow{\text{pose detector}} p_i(t) \quad \ \forall \ t = 1,\dots,T_i
\end{equation}
where $p_i(t)$ represents the $i$th sample pose at time (frame) $t$ and $T_i$ represents the time length of sample $i$.     
In particular, $p_i(t)$ consists of a list of 2D coordinates, namely: 
\begin{equation}
	p_i(t) = \{(x_j(t),y_j(t))_i\}_{j \in J}
\label{eq:rawsequences}
\end{equation} 
where $j$ represents the landmark index and $J$ is the landmarks set defined by the pose detector mapping. In this paper, $J = \{1,\dots,14\}$. Let $v_{j_1,j_2}$ be the link vector between body landmarks $j_1,j_2 \in J$ as defined in Fig. \ref{fig:humanpose}. 

ActionXPose aims to exploit RGB-poses mapping provided by the pose detector to generate informative time sequences to train a recurrent neural network using $\mathbb{T}$ to predict actions labels in $\mathbb{T}^*$.
\subsection{OpenPose-baseline}
\label{sec:baseline}
\begin{figure*}
\centering
\includegraphics[scale=0.35]{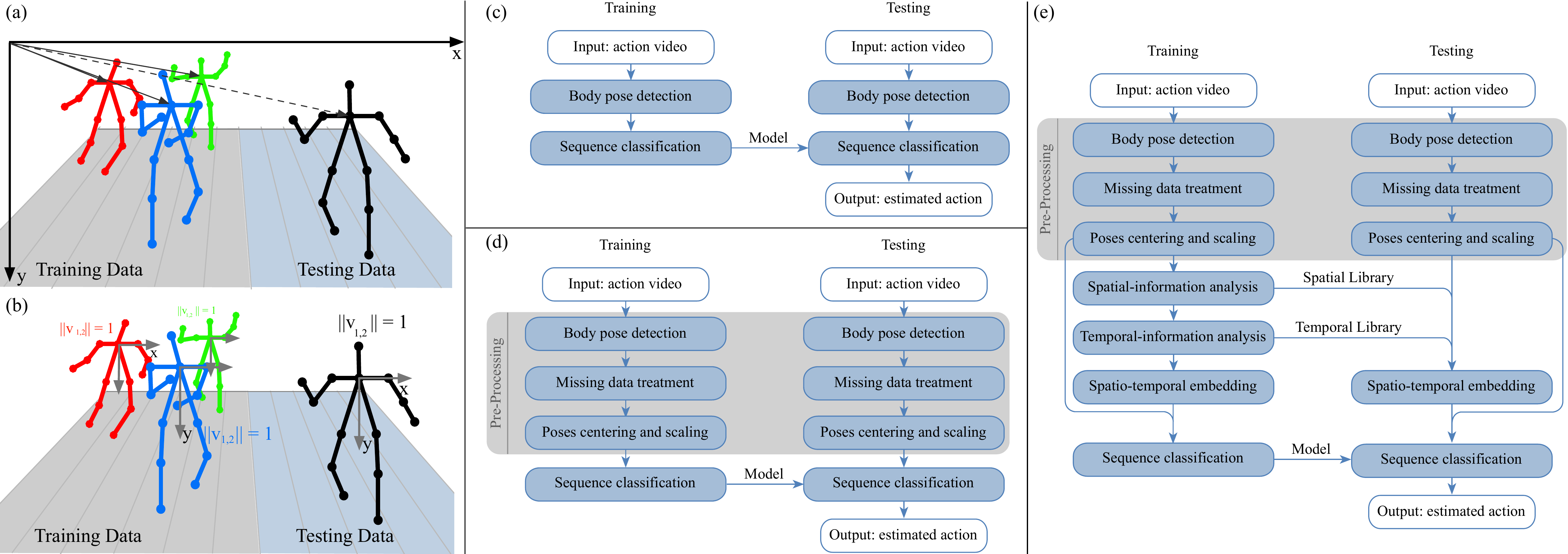}
\caption{(a) Since OpenPose provides landmarks coordinates in the global reference system, training data are location-dependent and size-dependent. (b) ActionXPose is based on centered and scaled poses, where each poses coordinates are given in a local reference system and pose size is transformed according to the constrain $||v_{1,2}|| = 1$. (c) \emph{OpenPose-baseline pipeline}. In the training phase, input videos are processed with OpenPose to extract body poses. Thus, landmark coordinates time sequences are provided to the sequence classification step to learn the temporal model. In the testing phase, learned model is exploited to classify testing sequences. (d) \emph{Proposed ActionXPose-basic pipeline}. In the training phase, input data is pre-processed to extract body poses, to treat missing data and to perform data centering and scaling. Thus, resulting body poses temporal sequences are exploited to train the model for the classification task. Such model is exploited in the testing phase, in order to get the final predicted label for testing data. (c) \emph{Proposed ActionXPose-advanced pipeline}. In the training phase, after input data pre-processing, resulting poses are clustered in the spatial and temporal analysis steps in order to create poses libraries. Then, spatio-temporal embedding is performed. Thus, embedded sequences alongside centered poses sequences train the classification model. In the testing phase, learned libraries are exploited to perform spatio-temporal embedding. Then, learned model is used to get the final prediction.}
\label{fig:general}
\end{figure*}

OpenPose \cite{Cao2016} is able to provide human poses represented by a root-centered graph made by 2D coordinates as shown in Fig. \ref{fig:humanpose}, where potentially some of them can be misdetected. As shown in Fig. \ref{fig:general}-(a), OpenPose coordinates locate each detected landmark into the processed frame according to a global coordinate system, where the origin of the axis is on the upper-left frame corner. As consequence, training and testing data are location and size dependent. This drawback motivates the proposed centering and scaling (Fig. \ref{fig:general}-(b)) strategy explained in Section \ref{sec:poses_centering}.
 
OpenPose-baseline consists of a simple learning step based on OpenPose coordinates sequences in (\ref{eq:rawsequences}). Since for some frames some body pose landmarks can be missing, OpenPose-baseline sets missing coordinates to an unexpected value, i.e. $-1$. This is a common solution for deep learning methods in presence of missing data \cite{GoodfellowIan2016Dl}. Regarding the learning model, Multivariate LSTM-FCN architecture with \emph{time-based attention mechanism} (MLSTM-FCN) \cite{Karim2018} is used, as in the ActionXPose cases. This algorithm takes as input landmarks coordinates in (\ref{eq:rawsequences}) obtained from training data $\mathbb{T}$, including action labels $l_i$, to train a supervised classification model. In the testing phase, this model takes as input landmarks sequences obtained from $\mathbb{T}^*$ and provides as output the estimated action label. In Fig. \ref{fig:general}-(c), the OpenPose-baseline pipeline is depicted. 

In Section \ref{sec:experiments}, OpenPose-baseline performance are provided as bottomline to measure ActionXPose performance and improvements. It turns out that location and size dependent training data compromise the generalisation performance when few training data are available, especially when testing actions are performed in a region of the frame not adequately covered by training data. This limitation motivates ActionXPose algorithm as evolution of OpenPose-baseline.

\section{ActionXPose}
\label{sec:ActionXPose}


\subsection{Overview}
\label{sec:overview}
In this paper, two ActionXPose variations are presented, namely ActionXPose-basic and ActionXPose-advanced.

\subsubsection{ActionXPose-basic} it is the simplest among the presented variations. In particular, it consists of only three cascade steps: 
\begin{enumerate}[i)]
	\item Missing data treatment (Section \ref{sec:missing_data_treatment});
	\item Poses centering and scaling (Section \ref{sec:poses_centering});
	\item Sequence classification (Section \ref{sec:sequences_classification}).
\end{enumerate}
In this method, pose data is treated in order to estimate missing landmarks. Secondly, pose is centered and scaled in order to make them consistent with each other, solving the OpenPose-baseline drawback. Finally, landmarks coordinates are converted into time-sequences to be provided to a sequence classifier. In Fig. \ref{fig:general}-(d), the proposed ActionXPose-basic pipeline for training and testing is depicted.

\subsubsection{ActionXPose-advanced} it consists of six cascade steps:
\begin{enumerate}[i)]
	\item Missing data treatment (Section \ref{sec:missing_data_treatment});
	\item Poses centering and scaling (Section \ref{sec:poses_centering});
	\item Spatial-information analysis (Section \ref{sec:spatio-information_analysis});
	\item Temporal-information analysis (Section \ref{sec:temporal-information_analysis});
	\item Gestures and Speed embedding (Section \ref{sec:spatio_temporal_embedding});
	\item Sequence classification (Section \ref{sec:sequences_classification}).
\end{enumerate}
The proposed ActionXPose-advanced is based on the same pre-processing and classification method steps as in ActionXPose-basic. However, it makes use of additional processing steps to extract full body and local posture related information. In Fig. \ref{fig:general}-(e), the proposed ActionXPose-advanced pipeline for training and testing is depicted. 

The main idea of ActionXPose is that, once human poses are extracted by the selected detector, the video-based HAR problem is reduced to a 2D pose-based HAR problem. Spatio-temporal action information is embedded frame-by-frame within the pose sequences. Thus, the strategy of both variations consists of learning spatio-temporal patterns from training data. While ActionXPose-basic relies only on pre-processed pose coordinates sequences, ActionXPose-advanced also considers additional sequences provided by two semi-supervised learning stages aimed to extract pose patterns from training data. Thus, such patterns are converted into time-sequences to be provided to the classification task. It turns out that ActionXPose-advanced is in general more accurate than ActionXPose-basic, although it requires more training time.


\subsection{Missing Data Treatment}
\label{sec:missing_data_treatment}
In this step, the problem of missing data is addressed as pre-processing step. Indeed, occlusions, self-occlusions or ambiguous RGB data can prevent the detector from detecting some body landmarks. In this paper, only missing data problem due to self-occlusions is addressed. Self-occlusions occur when the subject occludes itself assuming a particular posture in front of the camera. In general, poses can be affected by \emph{occasional missing data} which occurs in isolated frames, or \emph{persistent missing data} which occurs for the entire sequence.
Our strategy to deal with such missing data is structured in Algorithm \ref{missing_data_algorithm}.
The main ideas are:
\begin{itemize}
	\item If the root landmark is missing, then the target is not correctly detected. Thus, the pose is discarded;
	\item If more than 8 body landmarks are missing, the pose is assumed to be too poor to perform any HAR. Thus, the pose is discarded;
	\item If enough landmarks are available but some of them are missing, then we can approximate occasional missing data using spacial consistency, which means that missing landmarks are estimated from coordinates of the available landmarks;
	\item Persistent missing landmarks can be guessed by exploiting left and right side body symmetry;
\end{itemize} 
\begin{algorithm}
\caption{Missing data treatment algorithm.}
\label{missing_data_algorithm}
   \SetKwInOut{Input}{Input}
   \SetKwInOut{Output}{Output}
   \Input{2D poses with variable umber of missing landmarks}
   \Output{Refined 2D poses}
   \textbf{Initialization}: set time step $t = 0$\\
   \While{pose at time $t$ is available}{
   		$t = t + 1$\\
   		read pose at time $t$\\
   		\If{the root is missing {\bf or} number of missing landmarks $> 8$}{
   		skip current pose
  		}
   }
   \For{$i = 3$ to $14$}{
   		\uIf{$i$-th landmark is missing for all frames}{
   			replace $i$-th landmark data with the vertical specular landmark}
   			\Else{
   			interpolate $i$-th landmark missing data from available data in the time dimension \label{line:interpolation}
   		}
   	}
\end{algorithm}

This strategy is simple but effective. It is not intended to be the best estimation for missing data. Instead, it aims to be reasonable enough to support the subsequent processing.
The core of Algorithm \ref{missing_data_algorithm} is the interpolation step on line \ref{line:interpolation}, where occasional missing data is approximated by using spatial consistency. In formulations, we define $x(t)$ as the generic first coordinate of any landmark with respect to time $t$ where some entries are occasionally missing such that:
\begin{equation}
	x: A \longrightarrow \mathbb{R} \quad \quad A \subset \{1,\dots,T\}
\end{equation}
where $A$ represents the set of frames where the landmark is detected. Then, we define missing values for $t^* \in \{1,\dots,T\} \setminus A$ as the \emph{nearest-neighbour} as:
\begin{equation}
	x(t^*) = x\left(\hat{t}\right) \quad \text{s.t.} \quad \hat{t} = \operatorname*{arg\,min}_{t \in A} \lVert t^* - t \rVert_2
\end{equation}

It is worth mentioning that the output of Algorithm \ref{missing_data_algorithm} may still present \emph{persistent} missing data for landmarks such as $j\ne 2$, when even left and right symmetry fails, e.g. when both left and right landmarks are missing. This eventuality can be considered as due to a proper occlusion rather a self-occlusion (or possibly due to strong ambiguity between the target and the background). Further details on how to manage such cases are provided in Section \ref{sec:spatio_temporal_embedding}.  

\subsection{Poses Centering and Scaling}
\label{sec:poses_centering}
This step is intended to transform $p_i(t)$ from \emph{absolute} to the \emph{root-centered} coordinate reference system. The transformed pose coordinates are defined as:
\begin{equation}
(\bar{x}_j,\bar{y}_j)_i = (x_j,y_j)_i - (x_2,y_2)_i \quad \forall j \in J
\label{eq:translation}	
\end{equation}
where the dependence of $t$ has been conveniently omitted. Thus, 
\begin{equation}
\bar{p}_i=\{(\bar{x}_j,\bar{y}_j)_i\}_{j \in J}	
\end{equation}
is the set of root-centered coordinates defined by (\ref{eq:translation}).

Furthermore, ActionXPose defines 
\begin{equation}
\bar{\bar{p}}_i = \{(\bar{\bar{x}}_j,\bar{\bar{y}}_j)_i\}_{j \in J}
\label{eq:transformed_data}
\end{equation}
by scaling $\bar{p}_i(t)$ coordinates by using the following constrain
\begin{equation}
\bar{\bar{v}}_{j_1,j_2} = \frac{\bar{v}_{j_1,j_2}}{|| \bar{v}_{2,9} ||_2} \quad \forall j_1,j_2 \in J
\label{eq:scaling}
\end{equation}
where $\bar{v}_{2,9}$ is the vector link between the root and the right hip landmarks.

These two steps have two major advantages. First, they allow comparisons between poses reducing the effect due to target body size changes. Second, it makes the system robust to zooming, subject proximity and movements. In fact, (\ref{eq:translation}) conveniently centres pose roots over the origin of the axes. Thus, information about localization in the scene, i.e. absolute coordinates, are discarded. Moreover, exploiting (\ref{eq:scaling}), pose size become homogenous among different camera settings and target proximity. Thus, after this step, only information about the relative human body movements are kept for the further analysis.
\\With abuse of notation, for the rest of the paper, we will refer to $p_i(t)$ as the transformed poses in (\ref{eq:transformed_data}).

\begin{figure}
\centering
\includegraphics[scale=0.3]{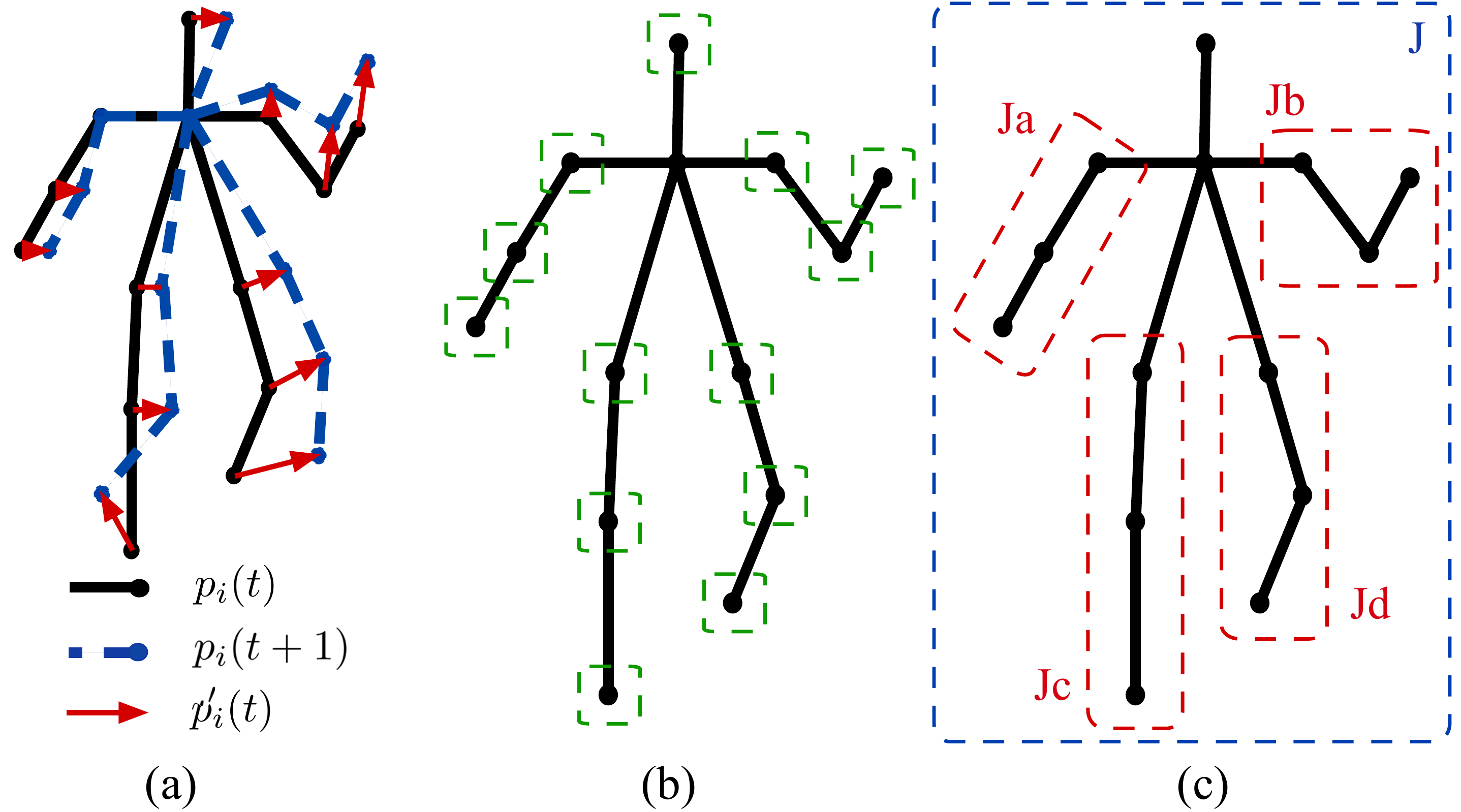}
\caption{(a) Human poses temporal information. Black pose represents $p_i(t)$. Blue pose represents $p_i(t+1)$. Red arrows represents $p'(t+1)$ and measures the displacements of $p_i(t+1)$ with respect to $p_i(t+1)$. (b) \emph{Sequence body attention}. $p_i(t)$ and $p_i'(t)$ carry information about single landmarks. Thus, ActionXPose-basic performs a landmark-level attention. (c) Subsets of landmarks for spatio-temporal embedding, where $J$ represents the entire set of landmarks. $J_a$ represents the right-arm landmarks set, while $J_b$ represents the left-arm landmarks set. Similarly, $J_c$ and $J_d$ represent, respectively, the right-leg and the left-leg landmarks sets. Thus, since ActionXPose-advanced make use of $p_i(t)$ and $p_i'(t)$ and $Seq_i(V_l)$ and $Seq_i(S_l)$, it allows different levels of body attention, namely landmark based, limbs based and full-body based.}
\label{fig:speeds_attention}
\end{figure}

As mentioned in Section \ref{sec:overview}, ActionXPose-basic only exploits landmark coordinates time-sequences $p_i(t)$ and $p_i'(t) = p_i(t+1)-p_i(t)$ to solve the recognition task. In particular, $p_i'(t)$ can be thought as a time sequence containing temporal information about the motion, while $p_i(t)$ only contain space information. In Fig. \ref{fig:speeds_attention}-(a), a graphic interpretation of $p_i'(t)$ is shown. Therefore, ActionXPose-basic allows a landmark-based attention (Fig. \ref{fig:speeds_attention}-(b)x). In other words, human action recognition is performed considering each landmark separately, to discriminate between different degrees of landmarks carried information. However, such level of details can be confusing in some cases, due to high intra-class variations. Therefore, in the subsequent sections, additional and meaningful time sequences will be defined, to be exploited alongside $p_i(t)$ and $p_i'(t)$ to enhance action recognition performance. Such additional sequences allow limbs-level and full body-level attention, to capture interesting local and general movements.

\subsection{Spatial Information Analysis}
\label{sec:spatio-information_analysis}
Since root coordinates in ${p}_i(t)$ have been set to zero in the previous section, let $u_i(t) = (x_1,y_1,x_3,y_3,\dots,x_J,y_J) \in \mathbb{R}^{2J-2}$ be the vector obtained by unrolling ${p}_i(t)$ and skipping the root coordinates $(x_2,y_2)_i$. Let also $\tilde{u}_i(t)$ be the vector containing the first $m$ principal component of $u_i(t)$ obtained by performing PCA. Thus, $\tilde{u}_i(t) \in \mathbb{R}^m$, where $m < 2J-2$. 

The main idea of this step is to exploit training data $\mathbb{T}$ to learn general poses for a fixed action and point of view that best represent the considered action. In other words, the output of this step will be a \emph{poses library} containing a certain number of pose prototypes per action and per point of view.

To this purpose, ActionXPose-advanced exploits Self-Organizing Map (SOM) \cite{kohonen-self-organizing-maps-2001} as unsupervised clustering method, to explore natural clusters within pose data into $\mathbb{R}^m$. Thus, for a fixed action label $l$ and a fixed point of view $w$, SOM is trained over 
\begin{equation}
\{\tilde{u}_i(t) \ | \ l_i = l, w_i = w \} \subset \mathbb{T}
\end{equation}
to define a homogenous topology of clusters $[q,\dots,q] \in \mathbb{R}^m$ where $q$ is a positive integer. This provides an additional cluster label $k_i$ for each training pose $\tilde{u}_i(t)$, as
\begin{align}
	&\{\tilde{u}_i(t) \ | \ l_i = l, w_i = w \} \xrightarrow{\text{SOM}} \nonumber \\
	\xrightarrow{\text{SOM}} &\{\tilde{u}_i(t) \ | \ l_i = l, w_i = w, k_i = k \} \quad \forall \ l \in \mathcal{L}
\end{align}
We recall that SOM runs over training data for fixed $l$ and $w$ as we are interested in exploring training data clusters action-by-action viewpoint-by-viewpoint. Thus, the output of SOM are cluster labels $k_i \in \{1,\dots,q^m\}$ provided for training data $\tilde{u}_i(t)$ for each action $l$, where $q^m$ is the maximum number of clusters defined by SOM architecture. We also mention that, since the SOM algorithm runs exploiting action labels $l$ and viewpoint labels $w$, this step can be considered as a semi-unsupervised learning step. 

Once training data have been clustered, $q^m$ pose prototypes are defined averaging clusters data as follows
\begin{align}
	\tilde{U}_{l,w,k}  = \frac{1}{n_k} \sum & \{\tilde{u}_i(t) \ | \ l_i = l, w_i = w, k_i = k \}\nonumber  \\ &\quad \forall \ k = 1, \dots, q^m
\end{align}
where $n_k$ represents the number of poses within cluster $k$.
In conclusion of this step, libraries of prototypes are collected from training data as follows
\begin{equation}
	V_l = \left\{ \{\tilde{U}_{l,1,k}\}_{k=1}^{q^m}, \dots, \{\tilde{U}_{l,\mathcal{W},k}\}_{k=1}^{q^m} \right\} \quad \forall \ l \in \mathcal{L}
\end{equation}
where we highlighted that for each action $l$, $V_l$ is obtained stacking all prototypes from different viewpoints.
Thus, $V_l$ contains pose prototypes in the form of points into a multidimensional space $\mathbb{R}^m$ able to cover all considered point of view variations. In other words, it contains \emph{peculiar} poses, representative of the training data for the entire action $l$ across different viewpoints.
Moreover, prototypes in $V_l$ are \emph{new} poses rather than some of the training poses. This choice has a major impact: it introduces more generality on the system, reducing overfitting in the overall learning process. 

It is worth mentioning that if training data are not carefully designed as common in real applications, $V_l$ can share poses. For example, if two different actions $l_1$ and $l_2$ start with the \emph{standing} position, both libraries $V_{l_1}$ and $V_{l_2}$ will contain prototypes for the standing position. We will discuss in Section \ref{sec:sequences_classification} how this affects the learning process.
   
\subsection{Temporal Information Analysis}
\label{sec:temporal-information_analysis}
In the previous section we have defined libraries for spatial-information. In this section we will discuss how to define temporal-information libraries.

Let $u_i(t) \in \mathbb{R}^{2J-2}$ be as defined in Section \ref{sec:spatio-information_analysis}. Therefore, temporal information resides in the mutual changes of $u_i(t)$ over time. Thus, we define $u_i'(t) = u_i(t) - u_i(t-1) \in \mathbb{R}^{2J-2}$.
This leads to the same analysis we discussed in Section \ref{sec:spatio-information_analysis}.
Thus, following the above mentioned steps, libraries for the temporal information remain defined as follows
\begin{equation}
	S_l = \left\{ \{\tilde{U'}_{l,1,k}\}_{k=1}^{q^m}, \dots, \{\tilde{U'}_{l,\mathcal{W},k}\}_{k=1}^{q^m} \right\} \quad \forall \ l \in \mathcal{L}
\end{equation}
where $\tilde{U'}_{l,w,k}$ represents prototypes obtained by clustering temporal vectors $u_i'(t)$.

\subsection{Spatio-temporal Embedding}
\label{sec:spatio_temporal_embedding}
In this section, spatio-temporal libraries $V_l$ and $S_l$ for $l \in \mathcal{L}$ are exploited to produce meaningful time sequences for each sample. The main idea is to measure frame-by-frame the \emph{minimum distance} between the detected pose with respect to the library poses. The lower is the minimum distance to the libraries for the action $l$, the more likely the pose on the current frame belongs to the action set $l$. Moreover, inspired by \cite{YongDu2015}, since different body parts could carry different information, we define a selective embedding procedure as follows.

Let $J_a,J_b,J_c,J_d$ be subsets of landmarks as defined in Fig. \ref{fig:speeds_attention}(b), namely
\begin{align}
\label{eq:setoflandmarks}
J &= \{1, \dots, 14\}\nonumber\\
	J_a &= \{3, 4, 5\} \subset J\nonumber\\
		J_b &= \{6, 7, 8\} \subset J\nonumber\\
			J_c &= \{9, 10, 11\} \subset J\nonumber\\
				J_d &= \{12, 13, 14\} \subset J
\end{align}

Thus, the averaged distance $d_{J_a}(p_i(t),v)$ between the generic pose $p_i(t)$ and the generic prototypes $v \in V_l$ computed considering landmarks $J_a$ is defined as follows
\begin{equation}
\label{eq:distances}
	d_{J_*}(p_i(t),v) = \frac{1}{\lvert J_* \rvert} \sum_{j\in J_*} \left\lVert (x_j,y_j)_i - \left(x^\dagger _j,y^\dagger _j\right)  \right\rVert _2
\end{equation}
where $(x^\dagger _j,y^\dagger _j)$ are coordinates of $j$-th landmark of $v$ and $J_*$ represents either $J_a$, $J_b$, $J_c$, $J_d$ or the whole set $J$.
Moreover, given a library of prototypes $V_l$ for action $l$, we can define the \emph{embedding sequence} as follows
\begin{equation}
	D_{V_l,J_*}(t) = \min_{v \in V_l} d_{J_*}(p_i(t),v)
\end{equation}
where it is clearly shown that $D_{V_l,J_*}$ dependents with respect to time.

Given a set of actions $\mathcal{L}$ and the set of landmarks in (\ref{eq:setoflandmarks}), the meaningful sequences that can be extracted from $p_i(t)$ remain defined as follows
\begin{align}
\label{eq:sequences}
	Seq_i(V_l) = &\{ D_{V_l,J}(t), D_{V_l,J_a}(t), \dots \nonumber \\
	&\dots, D_{V_l,J_b}(t), D_{V_l,J_c}(t), D_{V_l,J_d}(t) \} \quad \forall l \in \mathcal{L}
\end{align}

Equation (\ref{eq:sequences}) defines five sequences per action for each sample in the datasets. Thus, the full amount of sequences associated with the sample is $5\lvert \mathcal{L} \rvert$. The first sequence in (\ref{eq:sequences}) measures the similarity between the $i$-th sample with the prototypes in $V_l$. The second sequence measures the similarity between the right arm of the $i$-th sample with the right arm of the prototypes in $V_l$, and similarly for the other sequences.

Similarly, sequences for temporal information can be embedded as follows  
\begin{align}
\label{eq:temporal_sequences}
	Seq_i(S_l) = &\{ D_{S_l,J}(t), D_{S_l,J_a}(t), \dots \nonumber \\
	&\dots, D_{S_l,J_b}(t), D_{S_l,J_c}(t), D_{S_l,J_d}(t) \} \quad \forall l \in \mathcal{L}
\end{align}
where we simply replaced in (\ref{eq:sequences}) the term $V_l$ with $S_l$.
This leads to two set of sequences $Seq_i(V_l)$ and $Seq_i(S_l)$ for all $l \in \mathcal{L}$. 

It is worth mentioning that (\ref{eq:distances}) can be modified to cope \emph{persistent} missing data already mentioned in Section \ref{sec:missing_data_treatment}. In fact, few landmarks can be missing for the whole sequence, thus (\ref{eq:distances}) should be rewritten as
\begin{equation}
	d_{\bar{J}_*}(p_i(t),v) = \frac{1}{\left\lvert \bar{J}_* \right\rvert} \sum_{j\in \bar{J}_*} \left \lVert (x_j,y_j)_i - \left(x^\dagger _j,y^\dagger _j\right)  \right \rVert _2
\end{equation}  
where $\bar{J}_*$ denote $J_*$ without missing landmarks. This allow the system to run even in presence of persistent missing data. Moreover, since the distance $d_{\bar{J}_*}$ is defined as an averaged $L_2$-norm, it remains well-defined and consistent in all cases.


\begin{figure}
\centering
\includegraphics[scale=0.245]{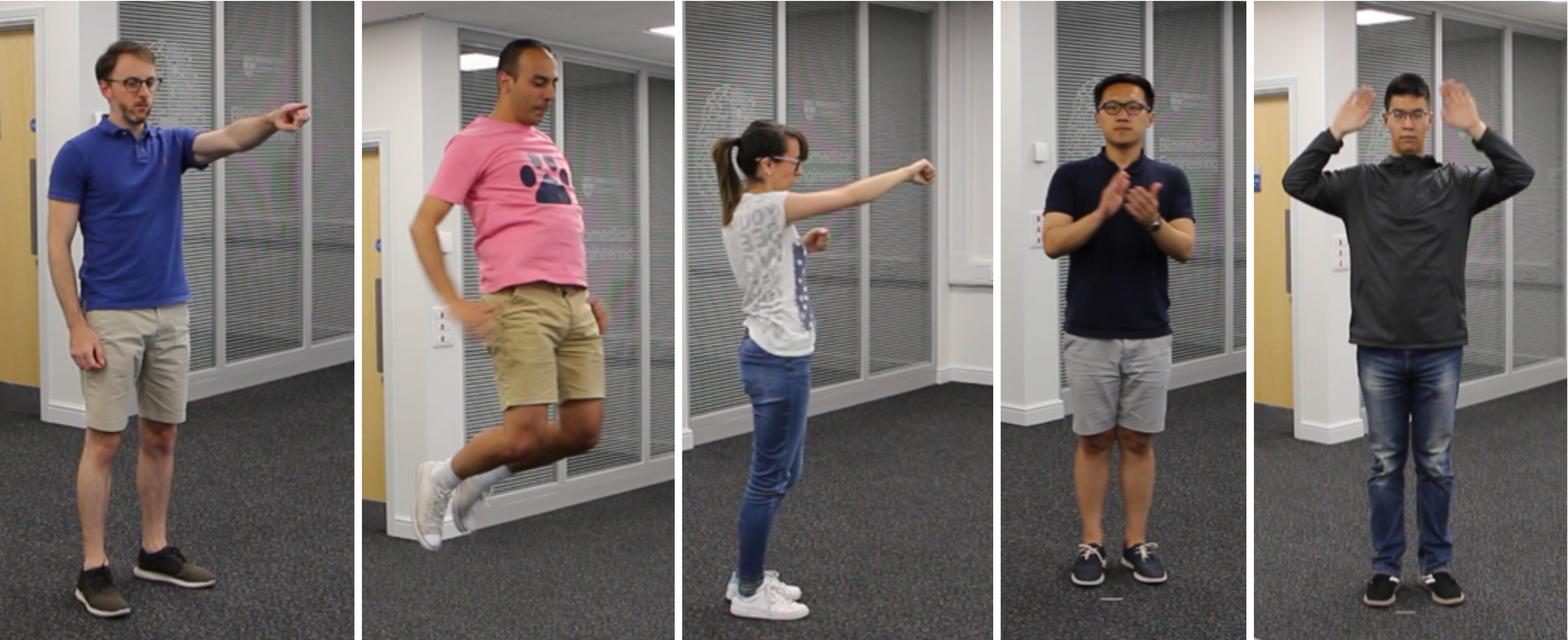}
\caption{Examples from ISLD dataset. Human actions of 10 subjects are recorded from a static camera from different viewpoints.}
\label{fig:isld}
\end{figure}

\subsection{Sequences Classification}
\label{sec:sequences_classification}
This Section aims to classify $p_i(t)$ and $p_i'(t)$ sequences defined in Section \ref{sec:poses_centering} alongside those defined by (\ref{eq:sequences}) and (\ref{eq:temporal_sequences}). Thus, the aim is to exploit training data $\mathbb{T}$ to generate a model for HAR.
ActionXPose addresses this problem by using the same model introduced in OpenPose-baseline, i.e. MLSTM-FCN \cite{Karim2018}. This model extracts convolutional and recurrent features from the sequences for time sequences based classification. 1D CNN as well as LSTM with time-attention are key parts of the model.
Therefore, according to the pipelines in Fig. \ref{fig:general}, ActionXPose-basic only provide $p_i(t)$ and $p_i'(t)$ as inputs to the MLSTM-FCN module, while ActionXPose-advanced provide the full list of sequences, namely $p_i(t)$, $p_i'(t)$, $Seq_i(V_l)$ and $Seq_i(S_l)$.
It is worth mentioning that, while ActioXPose-baseline only performs landmark-based attention, ActionXPose-advanced allows several attention levels, namely landmark-based, limbs-based and full-body based. Moreover, using the LSTM-FCN-Att sequence classifier, we are also able to consider time-based attention level. All these additional attention mechanisms make ActionXPose-advanced much more sophisticated than ActionXPose-basic, making it able to capture smaller details of the movement as well as local and full body postures. 

\section{MPOSE and ISLD datasets}
\label{sec:mpose_isld_dataset}

\subsection{Proposed MultiPose Dataset (MPOSE)}
\label{sec:proposed_multi_pose_dataset}
In this paper, we propose Multi Pose Dataset (MPOSE), which is a fusion between Weizmann, i3DPost, KTH and IXMAS samples. In particular, MPOSE contains only detected human poses, as original clips are publicly available. Regarding action labels, as shown in Table \ref{tab:datasets_actions}, they have been set in order to be consistent across the four datasets. Regarding action composition, we excluded from IXMAS actions \emph{throw}, \emph{throw-over-head}, \emph{throw-from-bottom-up} as in the original paper \cite{Weinland2006ixmas}. We also excluded \emph{turn-around} because in this case human poses temporal patterns are similar to those obtained by \emph{walking}. Moreover, we also excluded the viewpoint \emph{top} from IXMAS dataset because from that viewpoint the human body is barely visible. We also remark that some actions in Weizmann are peculiar of this dataset, such as \emph{jumping-jack} and \emph{gallop-sideways}. Clips for those actions are very limited in number in comparison with the other actions classes from other datasets, thus we excluded them from MPOSE to minimise classes imbalance.

Overall, MPOSE is a new and challenging datasets as it is made of video clips from four of the most popular datasets for HAR, providing detected human poses for 2D pose-based algorithm comparisons. In particular, it allows cross-datasets learning processing, as shown in Section \ref{sec:experiments}, which is one of the major goals of ActionXPose.

\subsection{Proposed Intelligent Sensing Lab Dataset (ISLD)}
\label{sec:isld}
\begin{figure}
\centering
\includegraphics[scale=0.12]{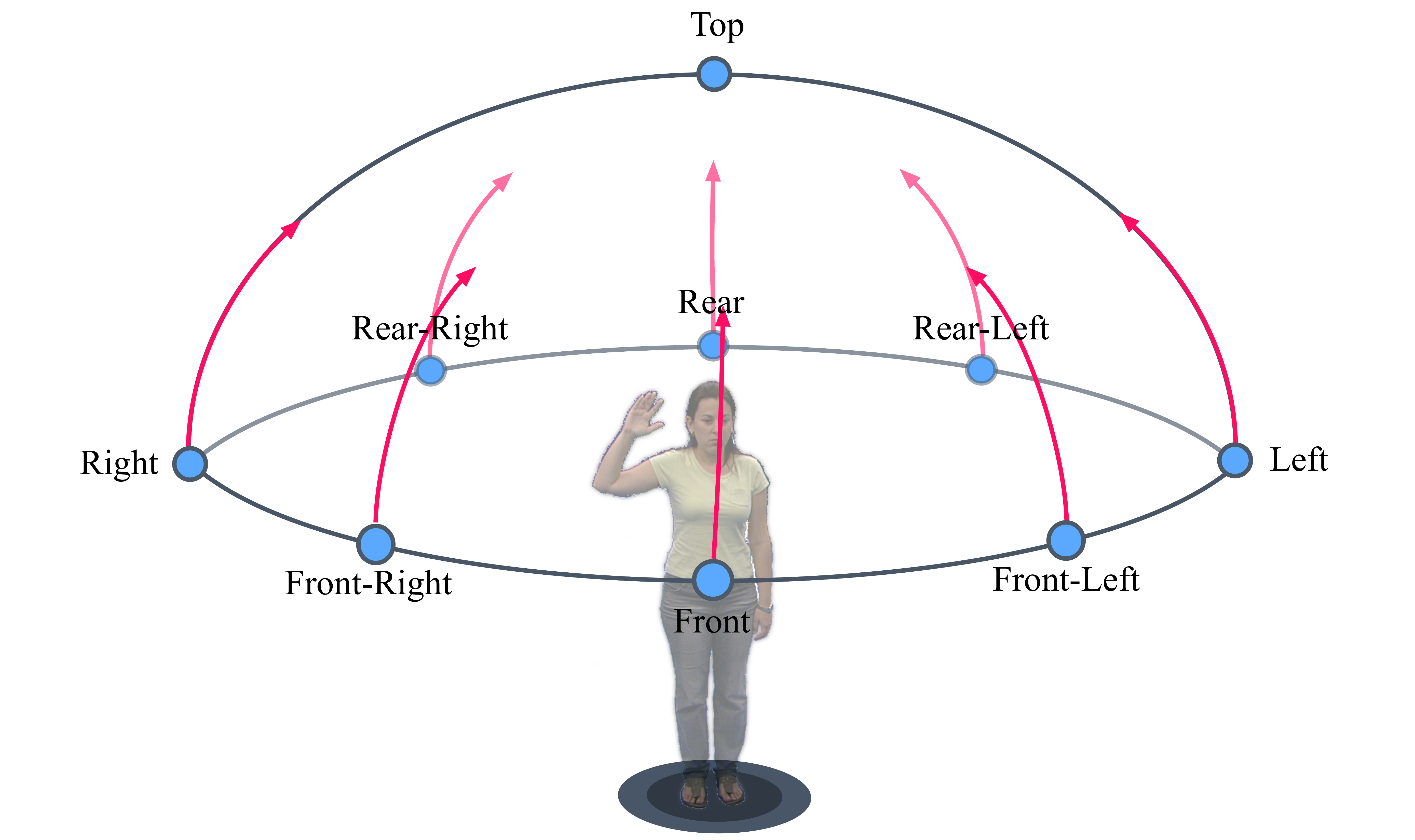}
\caption{Viewpoints nomenclature. The action can be recorded from one or many of these viewpoints, depending on the chosen dataset. Given by different recording environments, viewpoints can present vertical variability (pink arrows) across different datasets.}
\label{fig:pov}
\end{figure}

ISLD is a new dataset collected within our Intelligent Sensing Lab. It is inspired by the action composition of MPOSE, with the addition of the \emph{standing} action, which is a sort of \emph{no-action} interposed between each two actions. A total of 18 actions performed by 10 actors from 5 different viewpoints (\emph{front, front-left, front-right, left} and \emph{right}) have been recorded with a static RGB camera. Some frame examples of ISLD dataset are shown in Fig. \ref{fig:isld}. It is crucial to mention that no video examples from any of the previously mentioned datasets have been shown to ISLD participants. Before action recordings, participants have been only informed about action labels. Thus, they were free perform the actions according to their understanding of the action labels. Only recording viewpoints have been predefined. As a matter of fact, although MPOSE and ISLD share the same action labels (except for the \emph{standing} action), intra-class variation is remarkable. Moreover, although viewpoints are formally the same, highly vertical variations on the recording angle is still visible, as mentioned in Fig. \ref{fig:pov}. 

ISLD is provided with fixed training, validation and testing actors, namely
\begin{enumerate}
\item Training actors: $1, 2, 3$ and $4$;
\item Validation actors: $5$ and $6$;
\item Testing actors: $7, 8, 9$ and $10$.	
\end{enumerate}

The major novelty with respect to other datasets is that ISLD is suitable to measure transferable knowledge from MPOSE. In other words, by using ISLD testing samples, it is possible to measure how much knowledge from MPOSE is needed to cover real world scenario challenges, such as those proposed by ISLD. Cross-datasets experiments involving MPOSE and ISLD are reported in Section \ref{sec:experiments}. ISLD stand-alone tests are also reported in Section \ref{sec:experiments}. Action labels and considered viewpoints are summarised in Tables \ref{tab:datasets_actions} and \ref{tab:datasets_viewpoints}.

\section{Experiments}
\label{sec:experiments}
This Section provides extensive results for OpenPose-baseline, ActionXPose-basic and ActionXPose-advanced, including comparisons with the state-of-the-art.
 

It is worth mentioning that comparisons between ActionXPose and CNN based methods such as Two-stream are not possible at this stage due to limitations discussed in Section \ref{sec:introduction}. Moreover, any comparison between them will be not well defined, since ActionXPose is based on \emph{limited data}, i.e. body poses, while CNN based approaches fully exploits colour information, to extract features from both the target and the background. Hence, ActionXPose has been compared with the state-of-the-art methods on public datasets such as i3DPost and KTH, including comparisons with OpenPose-baseline. Moreover, testing on proposed new datasets MPOSE and ISLD are provided, to show generalisation performance.  

\begin{table}
\caption{Datasets action composition setting. Tested datasets action labels are reported, specifying considered (\cmark), not considered (\xmark) and not available (-) actions. To the purpose of this paper, some labels have been redefined to allow cross-datasets consistency.}
\centering
\addtolength{\tabcolsep}{-4pt}
\begin{tabular}{| c| c| c| c| c| c| c|}
\hline \rule{0pt}{2.3ex}
Action Label & Weizmann & i3DPost & KTH & IXMAS & MPOSE & ISLD\\ [0.5ex] 
\hline \rule{0pt}{2.3ex} 
bending 			& \cmark 		& \cmark  		&- 			&\cmark &\cmark & \cmark \\ 
boxing 				& - 			& -		  		&\cmark 	&\cmark &\cmark & \cmark \\
checking-watch		& - 			& -		  		& - 		&\cmark &\cmark & \cmark \\
crossing-arms		& - 			& -		  		& - 		&\cmark &\cmark & \cmark \\
gallop-sideways 		& \cmark 		& -			  	&- 			& -     &\xmark & - \\
getting-up			& - 			& -		  		& - 		&\cmark &\cmark & \cmark \\
hands-clapping 		& - 			& -			  	&\cmark 	& - 	&\cmark & \cmark \\
jogging 			& - 			& -			  	&\cmark 	& - 	&\cmark & \cmark \\
jumping 			& \cmark 		& \cmark  		&-  		& - 	&\cmark & \cmark \\
jumping-in-place		& \cmark 		& \cmark  		&- 			& - 	&\cmark & \cmark \\
jumping-jack 		& \cmark 		& -			  	&- 			& - 	&\xmark & - \\
kicking				& - 			& -		  		& - 		&\cmark &\cmark &\cmark \\
one-hand-waving 		& \cmark 		& \cmark		&-  		&\cmark &\cmark &\cmark \\
pointing			& - 			& -		  		& - 		&\cmark &\cmark &\cmark \\
running 			& \cmark		& \cmark 		&\cmark 	& - 	&\cmark &\cmark \\
scratching-head		& - 			& -		  		& - 		&\cmark &\cmark &\cmark \\
sitting-down			& - 			& -		  		& - 		&\cmark &\cmark &\cmark \\
skipping			& \cmark 		& -				& - 		& - 	&\xmark & - \\
standing			& -				& - 			& - 		& - 	& -		& \cmark \\
throw				& -				& -				& -		 	&\xmark &\xmark & - \\
throw-from-bottom-up& -				& -				& -		 	&\xmark &\xmark & - \\ 
turnaround			& -				& -				& -		 	&\xmark &\xmark & - \\
two-hands-waving 	& \cmark 		& -			  	&\cmark 	& - &\cmark &\cmark \\
walking 			& \cmark 		& \cmark		&\cmark 	&\cmark &\cmark &\cmark \\
\hline
\end{tabular}
\label{tab:datasets_actions}
\end{table}
\begin{table}
\caption{Datasets viewpoints composition. Tested datasets viewpoints are reported, specifying considered (\cmark), not considered (\xmark) and not available (-) viewpoints.}
\centering
\addtolength{\tabcolsep}{-2pt}
\begin{tabular}{| c| c| c| c| c| c| c |}
\hline \rule{0pt}{2.3ex}
 Viewpoints & Weizmann & i3DPost & KTH & IXMAS & MPOSE & ISLD\\ [0.5ex] 
\hline \rule{0pt}{2.3ex} 
front 			& \cmark 		& \cmark  	&\cmark & \cmark & \cmark & \cmark \\
front-left 		& - 				& \cmark  	&\cmark & \cmark & \cmark & \cmark \\
front-right 		& - 				& \cmark  	&\cmark & \cmark & \cmark & \cmark \\
left 			& \cmark 		& \cmark  	&\cmark & -  	 & \cmark & \cmark \\
right 			& \cmark 		& \cmark  	&\cmark & -  	 & \cmark & \cmark \\
rear 			& - 				& \cmark  	& -  	& -  	 & \cmark & -\\
rear-left 		& - 				& \cmark  	&\cmark & -  	 & \cmark & -\\
rear-right 		& - 				& \cmark  	&\cmark & -  	 & \cmark & -\\
top 			& - 				& -			& -  	& \xmark & -  	  & -\\ 
\hline 
\end{tabular}
\label{tab:datasets_viewpoints}
\end{table}

\subsection{OpenPose Setting}
\label{sec:openpose_setting}
OpenPose is a real-time multi-person 2D pose detector \cite{Cao2016}. It is designed to provide 18 body landmarks, 70 face landmarks, 42 hands landmarks and 6 feet landmarks for each target. To the purpose of this paper, only 18 body landmarks have been exploited. In particular, since 5 body landmarks represent \emph{nose}, \emph{left eye}, \emph{right eye} and \emph{left} and \emph{right ears}. In this paper, we defined a \emph{head} landmark averaging these landmarks, when available. Thus, in our implementation, the set of used body landmarks is $J = \{1,\dots,14\}$ as described in Figure \ref{fig:humanpose}.

We have used OpenPose version 1.2.1 (Demo for Windows). The used body model is based on COCO \cite{Lin2015} and we left OpenPose free to adjust the network resolution to the input video aspect ratio.
\subsection{Augmenting Training Data}
\label{sec:augmenting_training_data}
Deep learning methods usually requires great amount of data to perform as expected. However, many datasets for action recognition do not always contain enough labelled data. Usually, in fields such as Image Recognition, \emph{cropping} or \emph{rotating} images is common practice to augment dataset samples in order to meet deep learning algorithms conditions \cite{Jo2017}. In speech recognition, is also common to add noise to training samples to the same purpose \cite{Hsu2017}. In this paper, two strategies have been employed to augment training data: \emph{pose flipping} and \emph{pose noising}.

\subsubsection{Pose flipping}
it consists of flipping poses $p_i(t)$ for all $i\in \mathbb{T}$ along the vertical axis passing through the root landmark. This implies that the performed action looks flipped, exploiting left-right body symmetry. Moreover, resulting poses from the \emph{left} viewpoint are changed in the \emph{right} viewpoint and reverse. The same change occurs between \emph{front-left} and \emph{front-right}, as well as between \emph{rear-left} and \emph{rear-right}. Thus, such data augmentation provides additional variability to training data.\subsubsection{Pose noising} \label{sec:pose_noising} it consists of adding Gaussian noise to the $p_i$  landmark coordinates, for all $i\in \mathbb{T}$, i.e.
\begin{align}
	N(p_i) = 	\ [&(x_j,y_j)_i + (\tilde{x},\tilde{y})_1, \dots, \nonumber \\
				   &(x_j,y_j)_i + (\tilde{x},\tilde{y})_z]
\end{align}
where $\tilde{x}$ and $\tilde{y}$ are random real numbers generated by a gaussian distribution $\mathcal{N}(0,\sigma^2)$ with 0 mean and $\sigma$ standard deviation, and $z$ is the number of additional noised samples created with this procedure. Thus, if $z=0$, no noised samples are added to training data. If $z=1$, all training samples are used once to create additional training data. Similarly, if $z=2$, all training samples are used twice to create additional training data, and so on.
It is worth mentioning that, applying the pose noising $N$ to $p_i$ \emph{before} performing the embedding step (Section \ref{sec:spatio_temporal_embedding}) ensures that also the sequences $p_i'(t)$, $Seq_i(V_l)$ and $Seq_i(S_l)$ are noised. 

\begin{table}
\caption{Comparisons for KTH dataset performance. Accuracy results for Split and LOAO setting are reported.}
\centering	
\begin{tabular}{|c c c|} 
 \hline \rule{0pt}{2.3ex}
 Method & Setting & Accuracy\\[0.5ex]
 \hline
 \rule{0pt}{2.3ex}
 OpenPose-baseline & Split & 90.38\%\\
 OpenPose-baseline & LOAO & 95.21\%\\
 \hline
 \rule{0pt}{2.3ex}
 ActionXPose-basic & Split & 92.69\%\\
 ActionXPose-advanced & Split & 95.25\%\\
 ActionXPose-basic & LOAO & 98.91\%\\
 ActionXPose-advanced & LOAO & {\bf 99.04\%}\\
 \hline \rule{0pt}{2.3ex}
 Kovashka et al. \cite{Kovashka2010}& Split & 94.5\%\\
 Zhang et al. \cite{Zhang2012}& Split & 94.1\%\\
 Ji et al. \cite{ji2013}& Split & 90.2\%\\
 Selmi et al. \cite{Selmi2016}& Split & 95.8\%\\
 Liu et al. \cite{Liu2016a}& Split & 95\%\\
 Cheng-Bin Jin et al. \cite{Jin2017}& Split & {\bf 96.3\%}\\
 Almeida et al. \cite{Almeida2017}& LOAO & 98\%\\
 Vrigkas et al. \cite{Vrigkas2014}& LOAO & 98.3\%\\
 Liu et al. \cite{Liu2009}& LOAO & 93.8\%\\
 Raptis and Soatto \cite{Raptis2010}& LOAO & 94.5\%\\
 Jiang et al. \cite{Lin2012}& LOAO & 95.77\%\\
 Gilbert et al. \cite{Gilbert2011}& LOAO & 95.70\%\\
 \hline
\end{tabular}
\label{tab:kth_results}
\end{table}

\subsection{KTH Dataset Results}
\label{sec:kth_results}
In this section, ActionXPose results on KTH dataset are provided. The relevance of these tests is to allow comparisons between the proposed method and other methods. Moreover, since KTH contains multi-viewpoints samples, as well as challenging recording conditions such as camera moving, zooming in and out and different subjects proximities, these tests are also relevant to confirm ActionXPose claimed advantages. According to KTH literature, common setting for KTH tests are the following:
\subsubsection{Split setting}
in this setting, training and testing subjects are fixed as in the original paper \cite{Schuldt2004}. Since this test does not use cross-validation foldings, it does not explore the whole dataset variability. ActionXPose-basic and ActionXPose-advanced results have been summarised and compared with the state-of-the-art in Table \ref{tab:kth_results}.

\subsubsection{LOAO Setting}
Leave-one-actor-out (LOAO) cross-validation is a common setting to evaluate HAR performance on small datasets. It is intended to explore the whole dataset variability, maximising the gain from available data. Each cross-validation round tests over a particular subjects and trains the model over the rest of available subjects. The final accuracy is provided on average. For KTH dataset, 25 cross-validations foldings are required. ActionXPose results have been summarised and compared with state-of-the-art in Table \ref{tab:kth_results}. 

Hyper-parameters for both Split and LOAO settings were fixed for all tests. In particular, $z = 3$ and $\sigma = 3$.

ActionXPose-advanced outperforms both ActionXPose-basic and OpenPose-baseline in all settings. Moreover, in the Split setting, ActionXPose-advanced performance are among the state-of-the-art. In the case of LOAO setting, ActionXPose-advanced outperforms state-of-the-art methods.

\subsection{i3DPost Dataset Results}
\label{sec:i3DPost_results}
This dataset is usually tested with LOAO setting. It is specifically designed to cover cross-viewpoints action recognition or multi-camera action recognition. In this paper, we focus on the cross-viewpoints modality, as we are not interested in combining different point of views in stereo or multi-camera action recognition. Thus, results are given in the most difficult case in which the task is to recognise the target action by using data from one single camera in a multi-viewpoints mode. ActionXPose results are summarised in Table \ref{tab:i3dpost_results} and compared with the state-of-the-art. 

The relevance of this test is twofold: first, it allows comparisons with other methods; second, it shows that ActionXPose can effectively learn and recognise action from several viewpoints.

Also in this case, ActionXPose-advanced outperforms both ActionXPose-basic and OpenPose-baseline, achieving performance among the state-of-the-art.

\begin{table}
\caption{Accuracy results for i3DPost dataset (LOAO). $\lvert \mathcal{W} \rvert$ represents the number of considered viewpoints.}
\centering
\begin{tabular}{|c c c|} 
 \hline
 Method & $\lvert \mathcal{W} \rvert$ & Accuracy\\ [0.5ex] 
 \hline
 \rule{0pt}{2.3ex} 
 OpenPose-baseline & 8 & 91.66\%\\
 \hline
 \rule{0pt}{2.3ex} 
 ActionXPose-basic & 8 & 97.39\%\\
 ActionXPose-advanced & 8 & 98.95\%\\
 \hline
 \rule{0pt}{2.3ex} 
 Angelini et. al. \cite{Angelini2018icassp} & 8 & {\bf 99.73\%}\\ 
 Castro et al. \cite{Castro-Munoz2015} & 2 & 99.00\%\\ 
 Iosifidis et al. \cite{Iosifidis2013} & 8 & 98.16\%\\ 
 Azary et al. \cite{Azary2012} & 8 & 92.97\%\\
 Hilsenbeck et al. \cite{Hilsenbeck2016} & 8 & 92.42\%\\
 \hline
\end{tabular}
\label{tab:i3dpost_results}
\end{table}

\subsection{MPOSE Dataset Results}
\label{sec:mpose_results}
Generalisation degree for ActionXPose has been evaluated on MPOSE dataset. In this dataset, multi-viewpoint action samples are merged from four different datasets. Thus, the purpose of this section is to show cross-datasets performance. We recall that MPOSE dataset contains 17 actions recorded from 8 viewpoints, performed by 53 subjects in total.

To evaluate ActionXPose performance on MPOSE, we adopted the cross-validation setting, as used in many works in many different fields \cite{Castro-Munoz2015, Iosifidis2013, Azary2012, Hilsenbeck2016, Fei-Fei, Grauman2005, Berg2005, Kuhn2013}. LOAO setting is not suitable for this dataset for two major reasons. First, each subject perform only a subset of actions, thus LOAO setting would provide strongly unbalanced confusion matrices. Second, we would require 53 cross-validation rounds to complete a single LOAO test, which makes hyper-parameters tuning very intense and time-consuming. Thus, ActionXPose has been tested on MPOSE by using a 10 foldings Cross-Validation setting. In this setting, MPOSE samples have been randomly shuffled and split in 10 equal-size foldings. In particular, to allow the confusion matrix to be as fair as possible across different tests, we opted for an action-based splitting. Thus, samples from each action have been randomly shuffled and then split in 10 foldings. Therefore, the first folding of each action has been chosen for testing while the rest of them have been used for training. This strategy forces the number of tested samples per each action to be stable across different folding tests. The final accuracy result is provided on average over the all 10 possible choices of testing folding.

\begin{figure}
\centering
\includegraphics[scale=0.45]{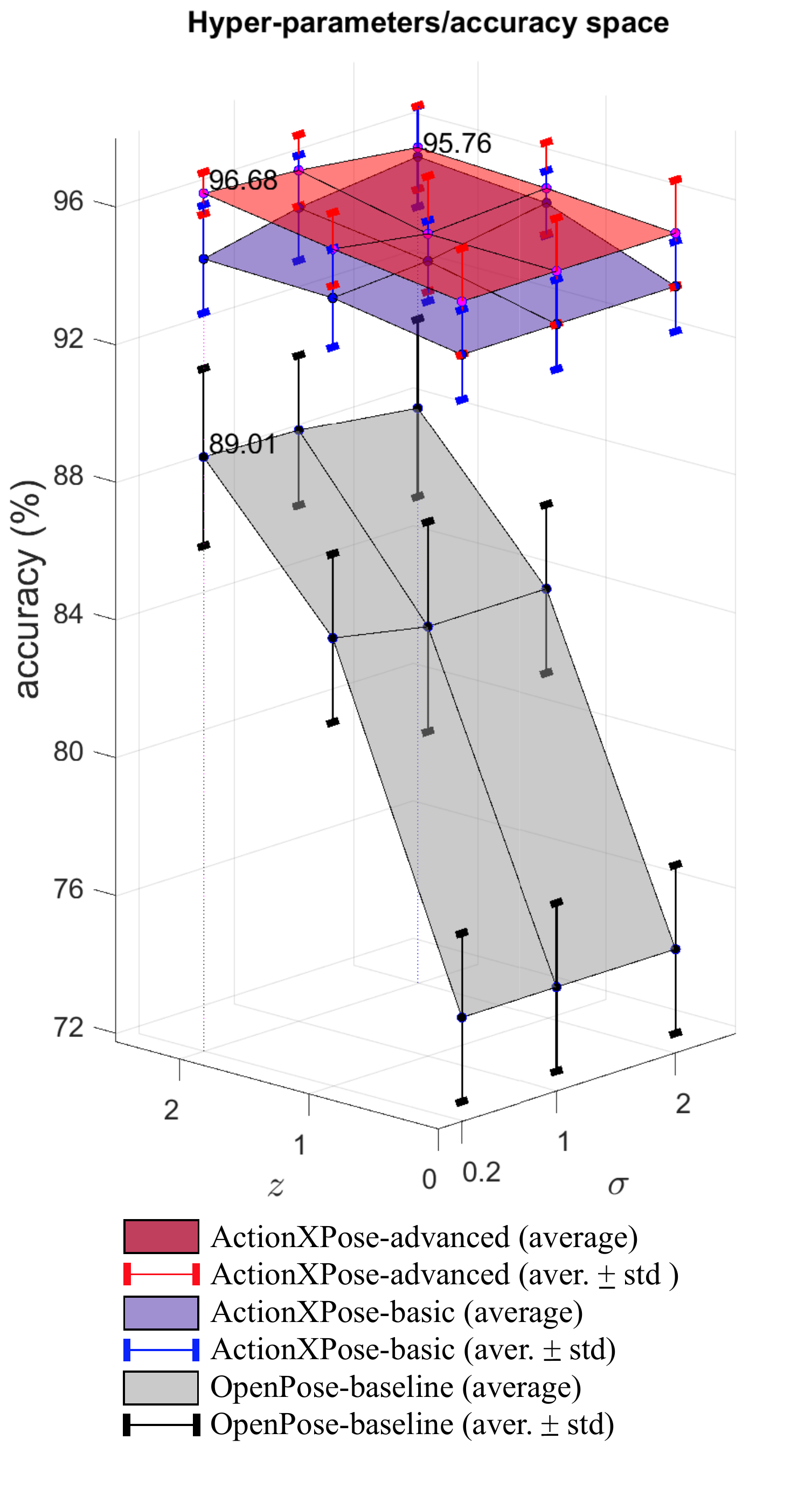}
\caption{MPOSE cross-validation accuracy results for different hyper-parameters settings, i.e. $\sigma$ and $z$. Pose flipping is always applied to training data. The graph reports averaged ed results for ActionXPose-advanced, ActionXPose-basic and OpenPose-baseline, including confidence intervals (average $\pm$ standard deviation). ActionXPose-advanced outperforms both ActionXPose-basic and OpenPose-baseline in all tests. The best setting for ActionXPose-advanced is with $z = 2$ and $\sigma = 0.2$, where the highest accuracy is reached with the minimum statistical variation.}
\label{fig:multipose_cross_settings}
\end{figure}

We mention that pose flipping and pose noising were applied to training samples. In particular parameters for pose noising, i.e. $z$ and $\sigma$ can be considered as hyper-parameters.
Results for the MPOSE test are shown in Fig. \ref{fig:multipose_cross_settings}, where we reported results for different hyper-parameter choices, i.e. $\sigma = \{0.2, 1, 2\}$ and $z = \{0, 1, 2\}$.

ActionXPose-advanced outperforms both ActionXPose-basic and OpenPose-baseline in all tests. As expected, performance for both methods increase when more training are provided, i.e. $z$ increases. Moreover, when the noise standard deviation increases, ActionXPose-basic performance increases accordingly. This is due to the fact that the noise disturbs the landmarks coordinates mutual correlation, forcing the network to generalize better, searching for rules that are preserved after noising. Conversely, ActionXPose-advanced seems not to benefit much from augmenting standard deviation noise.

Overall, these tests demonstrate that ActionXPose can effectively cover multi-datasets problems, showing high generalization degree. As a matter of fact, ActionXPose proves to be flexible enough to support a training phase where multiple datasets are provided, allowing tremendous potential when the number of provided datasets can be further increased.

\subsection{ISLD Dataset Results}
\label{sec:isld_results}
ISLD is novel dataset recorded in our Intelligent Sensing Lab. In this section, we only focus on ActionXPose-advanced as in the previous test it outperformed ActionXPose-basic and OpenPose-baseline. Tests are conducted on ISLD to measure algorithm performance in a common environment. However, ISLD is also particularly useful in this paper to measure MPOSE knowledge transferability. Thus, in this section, two sets of results are reported:
\begin{itemize}
\item Split setting: in this test, ActionXPose-advanced have been trained, validated and tested over predefined actor groups (see Section \ref{sec:isld}). It must be mentioned that the confusion matrix for this test is strongly unbalanced due to the presence of 149 \emph{standing} action clips, while the clips average for the other classes is around 8. Thus, absolute accuracy as well as relative accuracy are reported. 
\item ISLD-vs-MPOSE setting: in this test, MPOSE dataset has been used for training, while ISLD dataset has been used for validation and testing (according to predefined actor groups). Thus, in this test we evaluated how much MPOSE knowledge can be transferred to solve ISLD dataset. We remark that in this setting, as MPOSE do not contains data for the action \emph{standing}, we neglected ISLD validation and testing samples belonging to this class. Thus, the confusion matrix in this case is only slightly unbalanced. However, we provided absolute and relative accuracies for the sake of completnesses. 
\end{itemize}
Hyper-parameters for both settings were the same. In particular, $z = 0$, thus no pose noising were applied.  

\begin{figure*}
\centering
\includegraphics[scale=0.29]{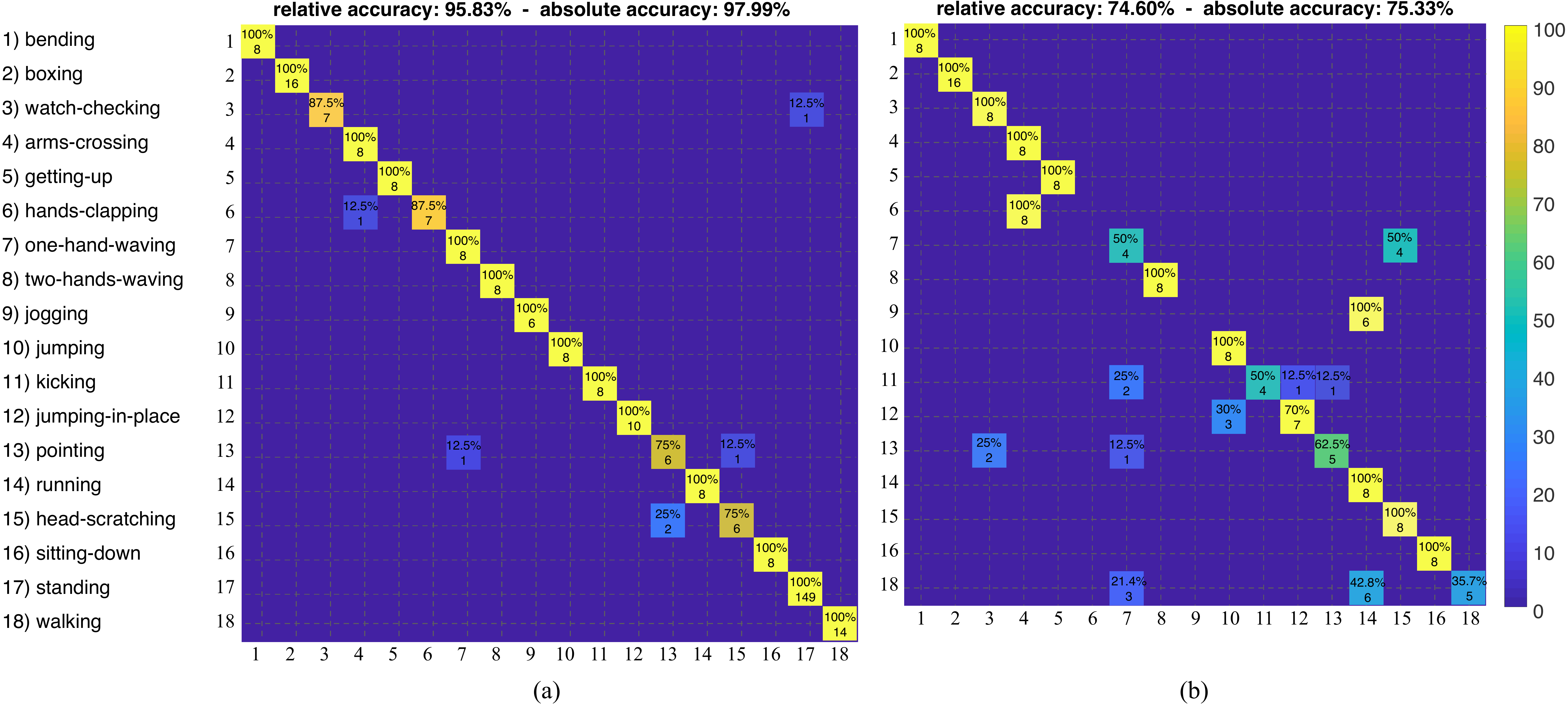}
\caption{ActionXPose-advanced results obtained on ISLD. Relative and absolute accuracies are shown. (a) ISLD Split setting confusion matrix. (b) ISLD Split setting confusion matrix.}
\label{fig:cm}
\end{figure*}

In the Split setting, ActionXPose-advanced achieved an absolute accuracy of 97.99\% and a relative accuracy of 95.83\% (Fig. \ref{fig:cm}-(a)). In the ISLD-vs-MPOSE setting, ActionXPose achieved an absolute accuracy of 75.33\% while a relative accuracy of 74.60\% (Fig. \ref{fig:cm}-(b)). This results reconfirm that while ActionXPose is still a good method for HAR achieving very good results on ISLD, MPOSE dataset knowledge is quite different from the knowledge included within ISLD dataset. Nevertheless, a substantial part of MPOSE can still train surprisingly well to address ISLD dataset action variability. This result also confirm once again how much human action recognition is an extremely difficult task due to the intra-class variations.

\subsection{Computational Speed Evaluation}
In this section, we provide computational speed evaluation for each of the most important processing steps required for ActionXPose. Our evaluation is made by considering separately \emph{training} and \emph{testing} phases. Speed evaluations have been obtained on Ubuntu 16.04 running on a laptop Dell Inspiron 15 5000 with 4 core Intel i7, mounting an embedded Nvidia GeForce GTX 1050.

Our simulations showed that the body pose detector is the \emph{bottleneck} for the entire processing. However, it is claimed to be a real-time detector \cite{Cao2016} when hardware requirements are fully satisfied.

Regarding the SOM clustering method, which is only required in the training phase, computational complexity can be estimated as $\mathcal{O}(N^2C^2)$ \cite{Roussinov1998}, where N is the number of considered samples and C the vector input dimensionality. This results shows the importance of considering PCA as dimensionality reduction technique before applying SOM. In ActionXPose, all conducted tests have been performed setting the PCA to consider only the first 3 principal components. Regarding used topology for SOM, 4 units per each component have been considered, which lead to pose libraries containing $4^3$ prototypes for each action and for each point of view. This relatively small number of prototypes is big enough to provide reported results. Moreover, it speeds up the subsequent embedding step for both training and testing phase. In fact, the embedding step has a complexity of $\mathcal{O}(n)$ where $n$ is the number of prototypes in the considered library \cite{Angelini2018icassp}. Thus, for small $n$ such as in our case, the embedding process is very fast. Our simulations shows that the embedding step can process a single frame in $\approx9*10^{-5}$ seconds, which corresponds to $\approx10^4$ frame per second. Even considering that ActionXPose needs to run two embedding processes (for the spatial and temporal libraries), the embedding process is still very fast. Overall, the full testing processing (excluding pose detection) for each clip sequence takes on average $4.1*10^{-2}$ seconds. Since in our tests each clip contains on average 72 frames, ActionXPose can elaborate its prediction with a speed of $5.7*10^{-4}$ seconds per frame. Thus, once hardware requirements for the pose detector are met, ActionXPose can run in real-time.

\subsection{Discussions}
\label{sec:results_discussion}

As we have seen in the previous sections, OpenPose-baseline is outperformed by both versions of ActionXPose, while ActionXPose-basic is in all cases less performing than ActionXPose-advanced. OpenPose-baseline failure cases are mostly due to the location and body size dependency of input data. ActionXPose-basic addresses this problem improving baseline results. The performance gap between ActionXPose-basic and ActionXPose-advanced is mostly due to the lower number of time-sequences that the basic version processes. Thus, ActionXPose-advanced is more accurate and effective for HAR problems, because it can rely on many meaningful sequences: single landmarks, limbs and full body sequences, both for spatial and temporal information.

The LSTM based sequence classifier allows to \emph{weight} these sequence singularly, providing natural framework for performing \emph{body-level attention}. Thus, during the training phase, the network can naturally learn which of the provided sequences are more meaningful, i.e. are carrying more information about the target action. Indeed, depending on the performed action, not all landmarks are equally informative as well as not all additional sequences regarding limbs and the full body do. Thus, performing body-level attention is a crucial feature for an effective HAR algorithm.

Moreover, the chosen classifier can also perform \emph{time-level attention}. Thus, during training, the network can learn which of the provided time steps are more informative. This additional level of learning is tremendously important, to allow great flexibility for the whole processing. In fact, ActionXPose formally does not require time-fixed-length inputs, due to the implemented LSTM framework. As a matter of fact, input sequences length can span a very high range of time steps. For example, in MPOSE datasets, data time lengths vary between 15 to 358 time-steps. This flexibility is one of the most important features for ActionXPose, because it allows interesting future work improvements regarding \emph{action detection} \cite{Singh2017}.

Regarding overall performance, ActionXPose achieves results among the state-of-the-art in all performed tests. In particular, even if in some cases is slightly under the best results in literature, it must be considered that ActionXPose uses \emph{limited data}, i.e. 14 body landmarks, to perform HAR. This suggests that ActionXPose is an effective method to process body posture information when multiple level of learning are required. In fact, by using ActionXPose for posture-related data extraction, CNN based additional levels of processing can explore data in different directions. 

Regarding our attempt to transfer knowledge from MPOSE to ISLD, achieved results suggest that ActionXPose can be further improved to try to generalise better in the training phase. In fact, when two very different datasets are combined, even though they share the same action labels, human attitude to the same action can be very different. This causes misclassifications that decrease performance. 

Overall, our tests show that ActionXPose can accept input data from several datasets at once, allowing cross-datasets learning processing. As consequence, we can train it to cover broad range actions, depending on the chosen applications.

In conclusion, ActionXPose achieved the following goals: 
\begin{enumerate}
	\item \emph{High generalisation degree}: MPOSE results show that ActionXPose can cope several datasets at once; in other words, it can effectively learn motion patterns from several sources in a common framework, allowing growth in performance simply adding data samples from different datasets;
	\item \emph{Multi-viewpoint robustness}: results shows great performance in dealing with several point of views at once; in fact, tested datasets contains actions recorded from up to 8 viewpoints;
	\item \emph{Low scenario restrictions}: this feature is due to the proposed pre-processing which is able to compensate different recording conditions, making pre-processed data consistent each other. Thus, the effects of the camera movings, zooming in and out as well as different subject proximities and movings on the scene are effectively compensated and removed from data; 
	\item \emph{State-of-the-art performance}: all tested results place ActionXPose performance among the state-of-the-art. It must be remarked that such good results have been achieved only using limited data, i.e. body landmarks coordinates;
	\item \emph{Real-Time performance}: ActionXPose processing run with very high frame-rate. The major hardware requirements is for the pose detector which in turn can run in real-time. 
\end{enumerate}

\section{Conclusions and Future Work}
\label{sec:conclusions}
In this paper, a novel human action recognition algorithm based on 2D human poses have been presented. ActionXPose is based on human pose detector to extract body landmarks for the target subject. Thus, ActionXPose provides pre-processing, extracts meaningful sequences from training data and learns body postures patterns for the classification task. LSTM and 1D CNN are exploited for the final learning and classification step.
ActionXPose showed very general learning abilities, being able to learn knowledge from single as well as from several different datasets at once. The proposed work also lays foundations for a deeper processing framework, which can also include RGB data alongside body landmarks.
In this paper, two novel datasets have been also introduced, MPOSE and ISLD. 

Future work will focus mainly on three directions. First, more generalisation ability is required to allow the system to work in real scenarios applications without additional training. Second, additional processing is needed to process background information as well as target related colour data. Thus, in future releases, ActionXPose will integrate CNN branches alongside the body-pose based processing proposed in this paper, making ActionXPose future releases to be able to address more challenging datasets such as ActivityNet and UCF101. Third, ActionXPose can be improved with automatic \emph{action detection} on online surveillance video sequences. Further work direction will be on defining ActionXPose-based \emph{anomaly detection} solutions where privacy-related restrictions do no allow colour data processing and storage.







%

\newpage
\bibliographystyle{IEEEtran}
\bibliography{../../../../MendeleyLib/library}

\end{document}